\documentclass[preprint,showpacs,preprintnumbers,amsmath,amssymb]{revtex4}
\usepackage{amsmath}

\usepackage{graphicx}
\usepackage{pdfpages}
\usepackage{dcolumn}
\usepackage{bm}
\usepackage{amsmath} 
\usepackage{amsfonts}
\usepackage{amssymb}
\usepackage{url}
\usepackage{microtype}
\usepackage{graphicx}
\usepackage{appendix}%

\newcommand{\qed}{\nobreak \ifvmode \relax \else
      \ifdim\lastskip<1.5em \hskip-\lastskip
      \hskip1.5em plus0em minus0.5em \fi \nobreak
      \vrule height0.75em width0.5em depth0.25em\fi}

\begin{document}

\preprint{}
\title{Archipelagos of Total Bound and Free Entanglement. II}
\author{Paul B. Slater}
 \email{slater@kitp.ucsb.edu}
\affiliation{%
Kavli Institute for Theoretical Physics, University of California, Santa Barbara, CA 93106-4030\\
}
\date{\today}
            
\begin{abstract}
In the indicated preceding preprint (I), we reported the results of, in particular interest here, certain three-parameter qubit-ququart ($2 \times 4$) and two-ququart ($4 \times 4$) analyses. In them, we relied upon entanglement constraints given by Li and Qiao. However, further studies of ours conclusively show--using the well-known necessary and sufficient conditions for positive-semidefiniteness that all leading minors (of separable components, in this context) be nonnegative--that certain of the constraints given are flawed and need to be replaced (by weaker ones). Doing so, leads to a new set of results, somewhat qualitatively different and, in certain respects, simpler in nature. For example, bound-entanglement probabilities of $\frac{2}{3} \left(\sqrt{2}-1\right)  \approx 0.276142$,  $\frac{1}{4} \left(3-2 \log ^2(2)-\log (4)\right) \approx 0.1632$, $\frac{1}{2}-\frac{2}{3 \pi ^2}  \approx 0.432453$ and $\frac{1}{6}$,  are reported for various implementations of constraints. We also adopt the Li-Qiao three-parameter framework to a two-parameter one, with interesting visual results.
\end{abstract}
 

\maketitle
Let us begin  by indicating the first model of a bipartite mixed state explicitly analyzed by Li and Qiao in their recent paper ``Separable Decompositions of Bipartite Mixed States" \cite{li2018separable}, and also in our  preceding preprint \cite{slater2020archipelagos}. It took the form 
 of the $2 \times 4$ dimensional mixed (qubit-ququart) state,
\begin{equation} \label{rhoAB}
\rho_{AB}^{(1)}=\frac{1}{2 \cdot 4} \textbf{1} \otimes \textbf{1} +\frac{1}{4} (t_1 \sigma_1 \otimes \lambda_1+t_2 \sigma_2 \otimes \lambda_{13}+t_3 \sigma_3 \otimes \lambda_3),
\end{equation}
where $t_{\mu} \neq 0$, $t_{\mu} \in \mathbb{R}$, and $\sigma_i$ and $\lambda_{\nu}$ are SU(2) (Pauli matrix) and SU(4) generators, respectively (cf. \cite{singh2019experimental}).

\begin{figure}
    \centering
    \includegraphics{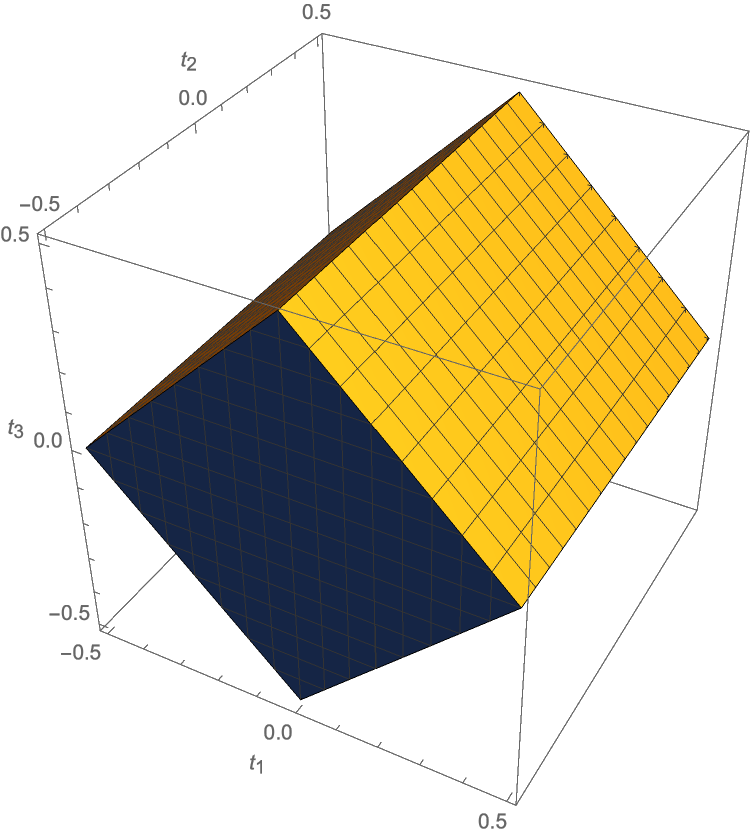}
    \caption{The convex set--in accordance with the constraints (\ref{eq2})--of possible qubit-ququart physical states representable by $\rho_{AB}^{(1)}$, given by (\ref{rhoAB}).}
    \label{fig:LiQiaoPPT}
\end{figure}

Li and Qiao found that equation (\ref{rhoAB}) represents a physical state when the $8 \times 8$ density matrix $\rho_{AB}^{(1)}$ is positive semidefinite, that is if
\begin{equation} \label{eq2}
t_2^2 \leq \frac{1}{4}, \hspace{.1in} (|t_1|+|t_3|)^2 \leq 
\frac{1}{4}.
\end{equation}
Figure~\ref{fig:LiQiaoPPT} shows the convex set of possible physical states representable by $\rho_{AB}^{(1)}$.
Let us now--to proceed in a probabilistic framework--standardize (dividing by one-half) the three-dimensional Euclidean volume of the possible physical states of $\rho_{AB}^{(1)}$  to equal 1.

Li and Qiao also established that  $\rho_{AB}^{(1)}$
 has positive (semidefinite) partial transposition, so the well-known PPT criterion could  not be used to help determine whether any specific state is entangled or separable.
Further, they asserted \cite[eq. (59)]{li2018separable} that $\rho_{AB}$ is entangled when \begin{equation} \label{eq3}
  (|t_1|+|t_2|+|t_3|)^2>1 \hspace{.1in}  \mbox{or} \hspace{.1in} (t_1 t_2 t_3)^2 > \frac{1}{27} \cdot \Big(\frac{2}{27}\Big)^2 =\frac{4}{27^3}=\frac{4}{19683} \approx 0.000203221,
\end{equation}
where they (correctly, we claim) associate the  quantity $\frac{1}{27}$ with the qubit and (incorrectly) the $\Big(\frac{2}{27}\Big)^2$ with the ququart.

Subsequent analyses of ours--using the well-known necessary and sufficient conditions for positive-semidefiniteness that all leading minors be nonnegative \cite{prussing1986principal}--firmly indicated that these constraints should be replaced by the decidedly weaker ones, 
\begin{equation} \label{EQ3}
  (|t_1|+|t_2|+|t_3|)^2> \frac{1}{2} \hspace{.1in}  \mbox{or} \hspace{.1in} (t_1 t_2 t_3)^2 > \frac{1}{27 \cdot 2^8} = \frac{1}{6912}  \approx 0.000144676.
\end{equation}
(We speculated that the $\Big(\frac{2}{27}\Big)^2$ bound could have, in fact, been obtained if some  different/nonstandard orderings--other than the one we employed, given in \cite[eq. (3)]{sbaih2013lie}--of the fifteen SU(4) generators had been employed. But for none of the possible three-member 455 subsets of the fifteen generators were such  bounds found. Interestingly, the discussion as to the variable ranges before eq. (67) in \cite{li2018separable} precisely agrees--using the relation $t_i =\alpha_i \beta_i$ and the bounds $\beta_1^2 +\beta_3^2 \leq \frac{1}{4}$, $\beta_2^2 \leq \frac{1}{4}$ and $\alpha_1^2+\alpha_2^2+\alpha_3^2 \leq 1$--with that we obtain using the leading-minors approach. However, the conclusions of Li and Qiao from these ranges are somewhat surprisingly incorrect--especially given their preceding detailed argument--as the maximization of $(|t_1|+|t_2|+|t_3|)^2$ and $(t_1 t_2 t_3)^2$ subject to the joint imposition of both  constraints yields $\frac{1}{2}$ and $\frac{1}{6912}$--and not 1 and $\frac{4}{19683}$, respectively.)

\begin{figure}
    \centering
    \includegraphics{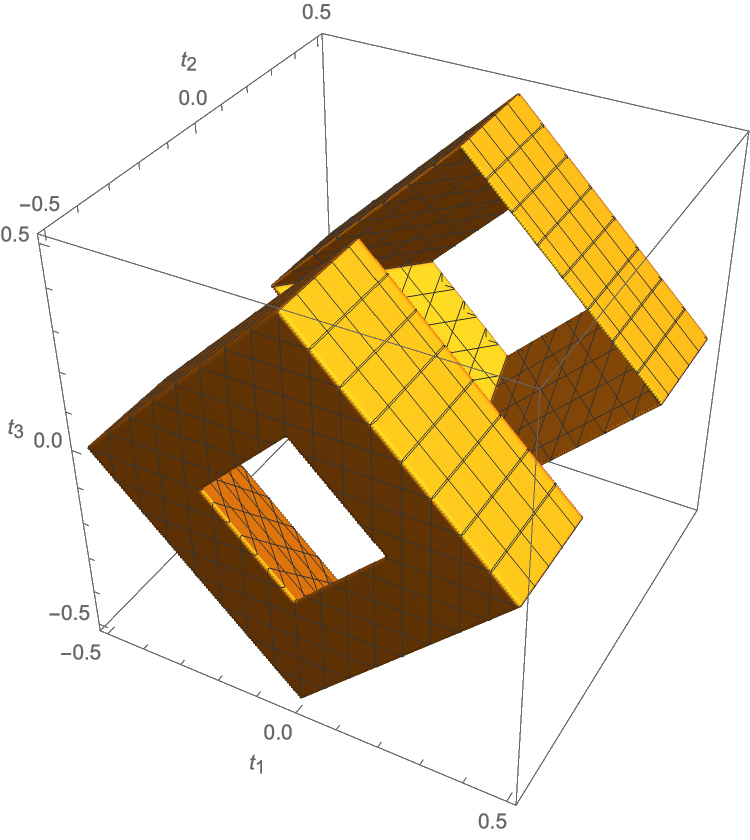}
    \caption{Qubit-ququart bound-entanglement islands--of probability $\frac{2}{3} \left(\sqrt{2}-1\right) \approx 0.276142$--given by enforcement  of the constraint $(|t_1|+|t_2|+|t_3|)^2> \frac{1}{2} $.}
    \label{fig:QubitQuquartAdditiveConstraint}
\end{figure}

This replacement of entanglement bounds immediately leads us to a remarkable result. While the constraint 
$(|t_1|+|t_2|+|t_3|)^2>1 \hspace{.1in}$ given by Li and Qiao proved to be unenforceable/irrelevant (perhaps an indication of its incorrectness), the weaker constraint $(|t_1|+|t_2|+|t_3|)^2> \frac{1}{2} $ gives us a bound-entanglement qubit-ququart probability of $\frac{2}{3} \left(\sqrt{2}-1\right) \approx 0.276142$ and an accompanying pair (archipelago) of corresponding islands (Fig.~\ref{fig:QubitQuquartAdditiveConstraint}).

\begin{figure}
    \centering
    \includegraphics{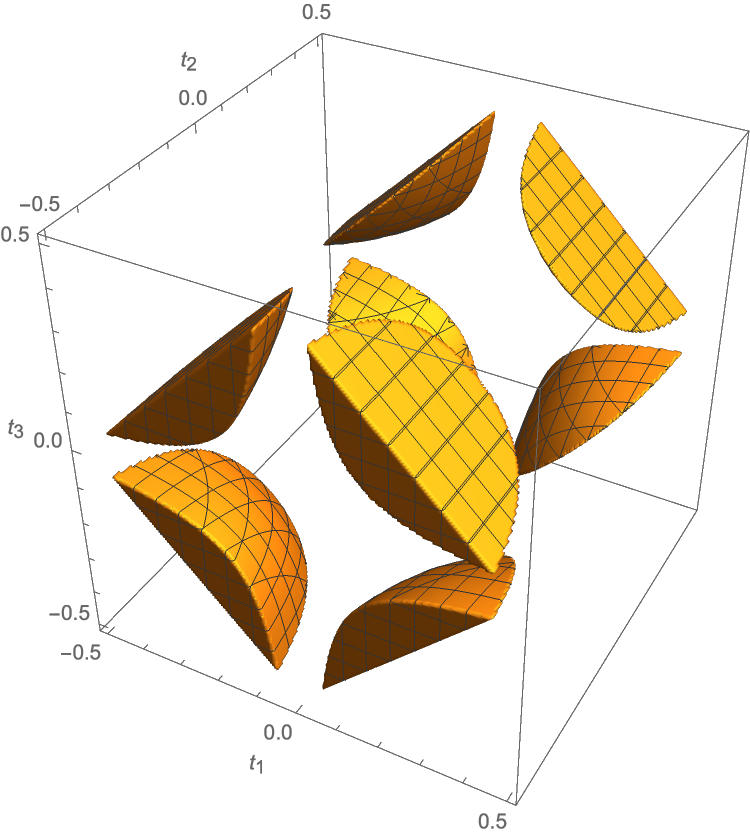}
    \caption{Qubit-ququart bound-entanglement islands--of probability $ \approx 0.12668688797$--given by enforcement  of the constraint $(t_1 t_2 t_3)^2 > \frac{1}{6912}$.}
    \label{fig:QubitQuquartMultiplicativeConstraint},
\end{figure}

Use of the further (multiplicative) constraint $(t_1 t_2 t_3)^2 >  \frac{1}{6912}$ gives us a more complicated (and smaller) bound-entanglement probability ($\approx 0.12668688797$) of 
\begin{equation} \label{QubitQuquartMultiplicative}
\frac{p}{3}+\frac{\log (12) \log (419904)}{48 \sqrt{3}}+\frac{\log (3)}{6
   \sqrt{3}}+\frac{\log (2)}{3 \sqrt{3}}
\end{equation},
\begin{displaymath}
-\frac{\log (18) \log (p+3)}{6 \sqrt{3}}-\frac{2 \log (p+3)}{3 \sqrt{3}}-\frac{\log (6)
   \log \left(12 \left(\sqrt{3} p+3 \sqrt{3}-1\right)\right)}{12 \sqrt{3}}
\end{displaymath}
\begin{displaymath}
+\frac{\text{Li}_2\left(\frac{p+3}{6}\right)}{3
   \sqrt{3}}-\frac{\text{Li}_2\left(\frac{3-p}{6}\right)}{3 \sqrt{3}},
\end{displaymath}
where $p=\sqrt{9-2 \sqrt{3}} \approx 2.35285$, and the polylogarithmic (dilogarithmic) function is employed. The corresponding archipelago diagram is Fig.~\ref{fig:QubitQuquartMultiplicativeConstraint}.

\begin{figure}
    \centering
    \includegraphics{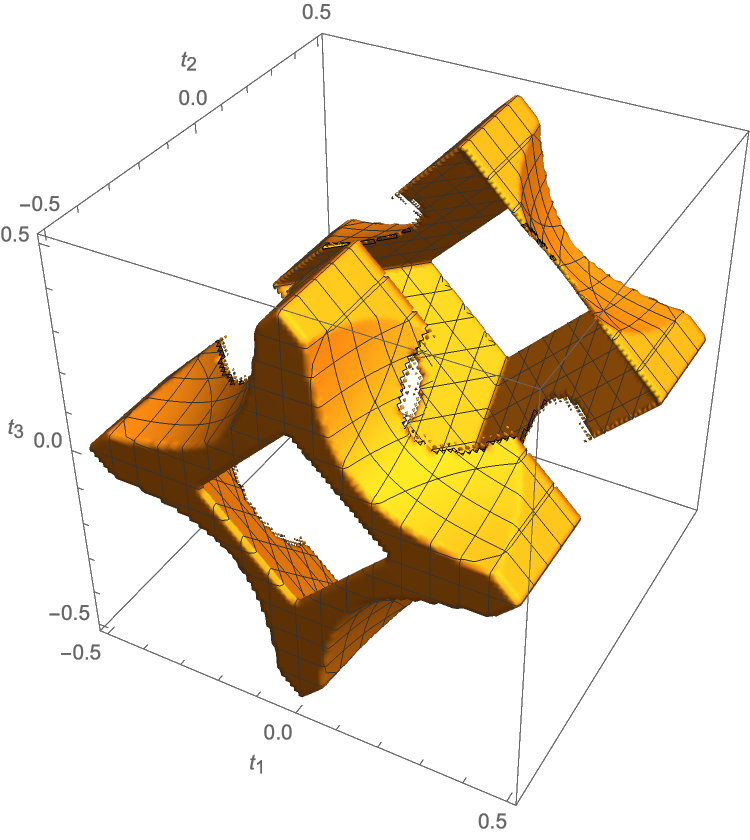}
    \caption{Bound-entangled archipelago of those qubit-ququart states satisfying the constraint $(|t_1|+|t_2|+|t_3|)^2> \frac{1}{2} $, but not $(t_1 t_2 t_3)^2 > \frac{1}{6912}$. The associated probability is approximately 0.151609.}
    \label{fig:QubitQuartOneConstraint}
\end{figure}

In Figure~\ref{fig:QubitQuartOneConstraint}, we show the bound-entangled  archipelago of those qubit-ququart states satisfying the constraint $(|t_1|+|t_2|+|t_3|)^2> \frac{1}{2} $, but now not $(t_1 t_2 t_3)^2 > \frac{1}{6912}$.
The associated probability is approximately 0.151609 \cite{ImplicitVolume}.

\begin{figure}
    \centering
    \includegraphics{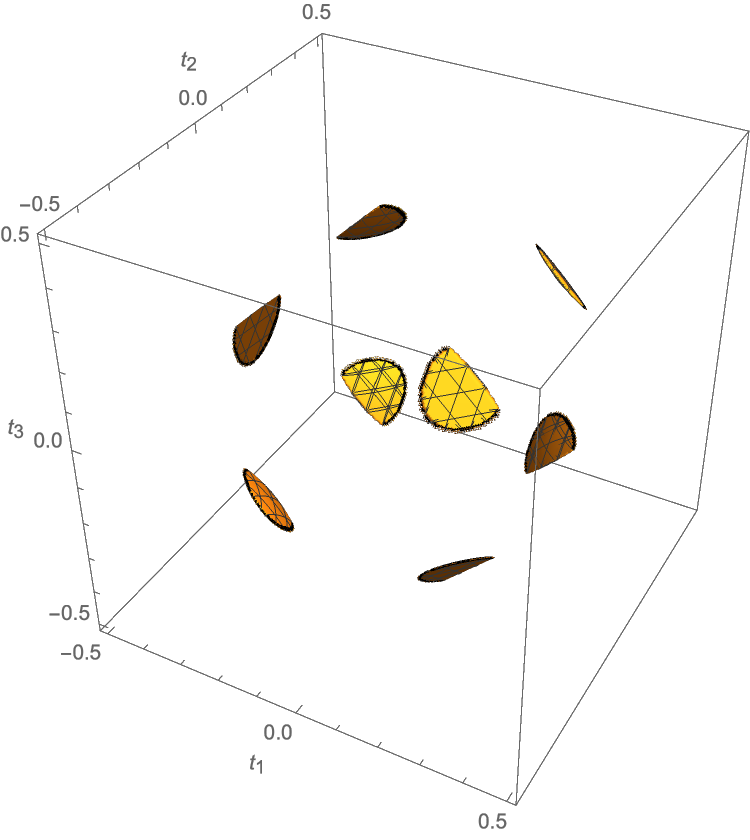}
    \caption{Bound-entangled archipelago of those qubit-ququart states satisfying the constraint $(t_1 t_2 t_3)^2 > \frac{1}{6912}$, but not  $(|t_1|+|t_2|+|t_3|)^2> \frac{1}{2} $. The associated probability is quite negligible, that is approximately 0.000269161439.}
    \label{fig:QubitQuartOneConstraint2}
\end{figure}

Reversing matters, in Figure~\ref{fig:QubitQuartOneConstraint2}, we show the bound-entangled  archipelago of those qubit-ququart states  satisfying the constraint $(t_1 t_2 t_3)^2 > \frac{1}{6912}$ but not  $(|t_1|+|t_2|+|t_3|)^2> \frac{1}{2} $. The associated probability is quite negligible, that is approximately 0.000269161439.

The probability that both (additive and mulitiplicative-type) constraints are satisfied is approximately 0.11265766, and the (total bound) probability that at least one of the two constraints is satisfied is approximately 0.276411536. (An accompanying plot for the first probability appears as a somewhat diminished version of Fig.~\ref{fig:QubitQuquartMultiplicativeConstraint} and an accompanying plot for the second probability appears a somewhat expanded version of Fig.~\ref{fig:QubitQuquartAdditiveConstraint}.)

\begin{figure}
    \centering
    \includegraphics{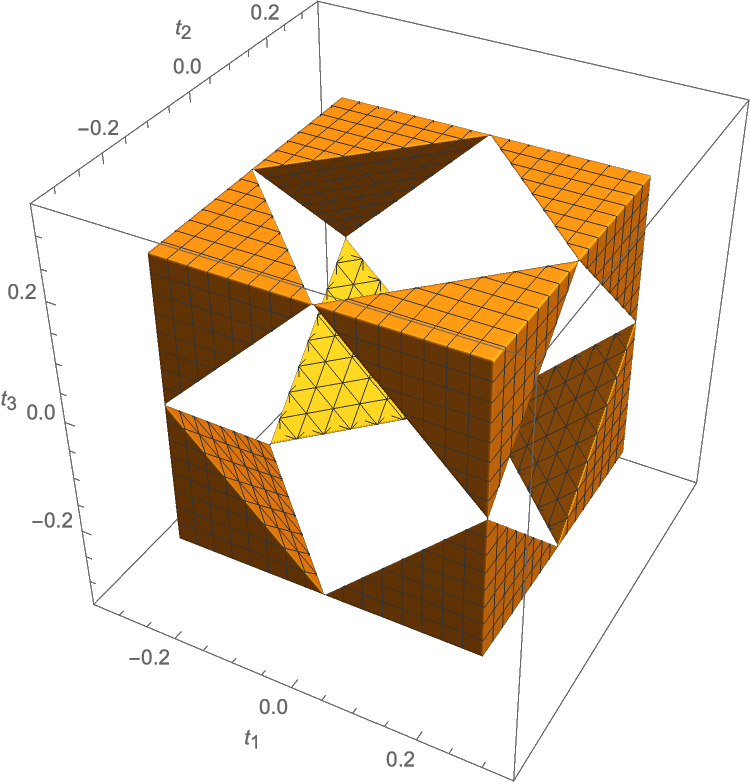}
    \caption{Two-ququart (\ref{2ququarts}) bound-entanglement islands--of probability $\frac{1}{6} \approx 0.166666$ --given by enforcement  of the constraint $(|t_1|+|t_2|+|t_3|)^2> \frac{1}{4} $.}
    \label{fig:TwoQuquartAdditiveConstraint}
\end{figure}

\begin{figure}
    \centering
    \includegraphics{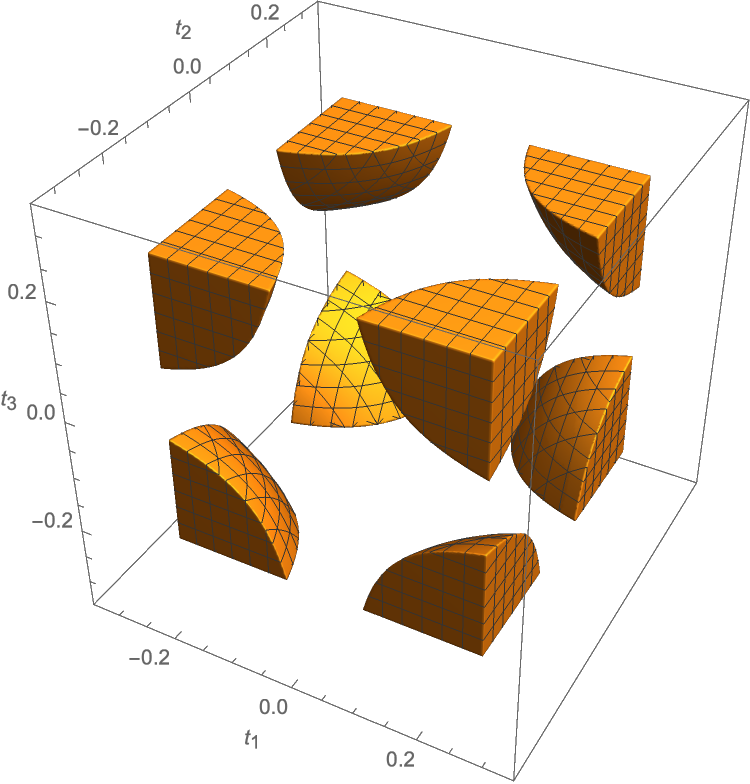}
    \caption{Two-ququart (\ref{2ququarts}) bound-entanglement islands--of probability $\frac{1}{4} \left(3-2 \log ^2(2)-\log (4)\right) \approx 0.1632$ --given by enforcement  of the constraint $(t_1 t_2 t_3)^2 >  \frac{1}{65536} $.}
    \label{fig:TwoQuquartMultiplicativeConstraint}
\end{figure}

Let us now shift--as we had in \cite{slater2020archipelagos}--to the study of the two-ququart states,
\begin{equation} \label{2ququarts}
\rho_{AB}^{(2)}=\frac{1}{2 \cdot 8} \textbf{1} \otimes \textbf{1} +\frac{1}{4} (t_1 \lambda_1 \otimes \lambda_1+t_2 \lambda_{13} \otimes \lambda_{13}+t_3 \lambda_3 \otimes \lambda_3),
\end{equation}
where as before the $\lambda$'s are $SU(4)$ generators.
The set of all two-ququart states is delimited by the constraint
\begin{equation} \label{twoququarts}
 -\frac{1}{4}<t_2<\frac{1}{4}\land -\frac{1}{4}<t_1<\frac{1}{4}\land
   -\frac{1}{4}<t_3<\frac{1}{4}.  
\end{equation}
That is, the set of possible $\{t_1,t_2,t_3\}$ comprises the cube $[-\frac{1}{4},\frac{1}{4}]^3$.
All these states have positive partial transposes, so all entangled states are  bound.
 Then, we have-again using the well-known necessary and sufficient conditions for positive-semidefiniteness that all leading minors be nonnegative \cite{prussing1986principal}-the corresponding 
 entanglement constraints (cf. eq. (\ref{eq3})),
\begin{equation} \label{EQ4}
  (|t_1|+|t_2|+|t_3|)^2> \frac{1}{4} \hspace{.1in}  \mbox{or} \hspace{.1in} (t_1 t_2 t_3)^2 > \frac{1}{2^{16}} = \frac{1}{65536}  \approx 0.0000152588,
\end{equation}
rather than $\frac{1}{2}$ and $\frac{1}{6912}$ as in the qubit-ququart model. (Again, we note for our maximization procedures, the basic relation in the Li-Qiao framework, $t_i =\alpha_i \beta_i$, together with the bounds,   $\alpha_1^2 +\alpha_3^2 \leq \frac{1}{4}$, $\alpha_2^2 \leq \frac{1}{4}$     and    $\beta_1^2 +\beta_3^2 \leq \frac{1}{4}$, $\beta_2^2 \leq \frac{1}{4}$ .) 
The single constraint $(|t_1|+|t_2|+|t_3|)^2> \frac{1}{4}$ gives us a bound-entanglement probability of $\frac{1}{6} \approx 0.166666$ and a set of corresponding islands (Fig.~\ref{fig:TwoQuquartAdditiveConstraint}).
The single constraint $(t_1 t_2 t_3)^2 >  \frac{1}{65536} $ yields a roughly equal-sized  bound-entanglement probability of $\frac{1}{4} \left(3-2 \log ^2(2)-\log (4)\right) \approx 0.1632$, and a set of corresponding islands (Fig.~\ref{fig:TwoQuquartMultiplicativeConstraint}).

The probability that both (additive and mulitiplicative) constraints are satisfied is approximately 0.149164132389, while 
the (total bound) probability that either of the two constraints is satisfied is 
approximately 0.180702437039. These two probabilities, of course, add to $\frac{1}{6} +\frac{1}{4} \left(3-2 \log ^2(2)-\log (4)\right)=\frac{1}{12} \left(11-6 \log ^2(2)-3 \log (4)\right) \approx 0.3298665694275933$ (as a matter of Boolean
logic, since $(A\land B)\lor (A\lor B)=A\lor B$).
 Our best current efforts at exactly computing this pair of  probabilities yielded  expressions employing elliptic integrals
plus one-dimensional integrals of $t_1$ over $[-\frac{1}{4}, \frac{1}{8}(\sqrt{5}-3)]$ and over $[\frac{1}{8} (3-\sqrt{5}),\frac{1}{4}]$, many of the integrands involving the term  $\sqrt{64 \left(t_1-1\right) t_1-\frac{1}{t_1}+16}$.

\begin{figure}
    \centering
    \includegraphics{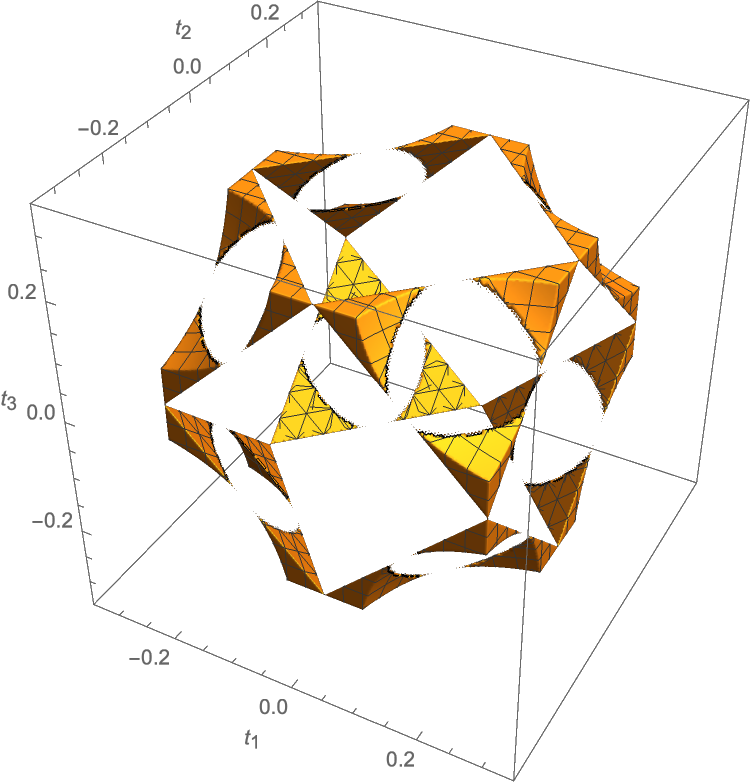}
    \caption{Bound-entangled archipelago of those two-ququart states (\ref{2ququarts}) satisfying the constraint $(|t_1|+|t_2|+|t_3|)^2> \frac{1}{4} $, but not $(t_1 t_2 t_3)^2 > \frac{1}{65536}$. The associated probability is approximately 0.0175025342.}
    \label{fig:TwoQuquarttOneConstraint}
\end{figure}

\begin{figure}
    \centering
    \includegraphics{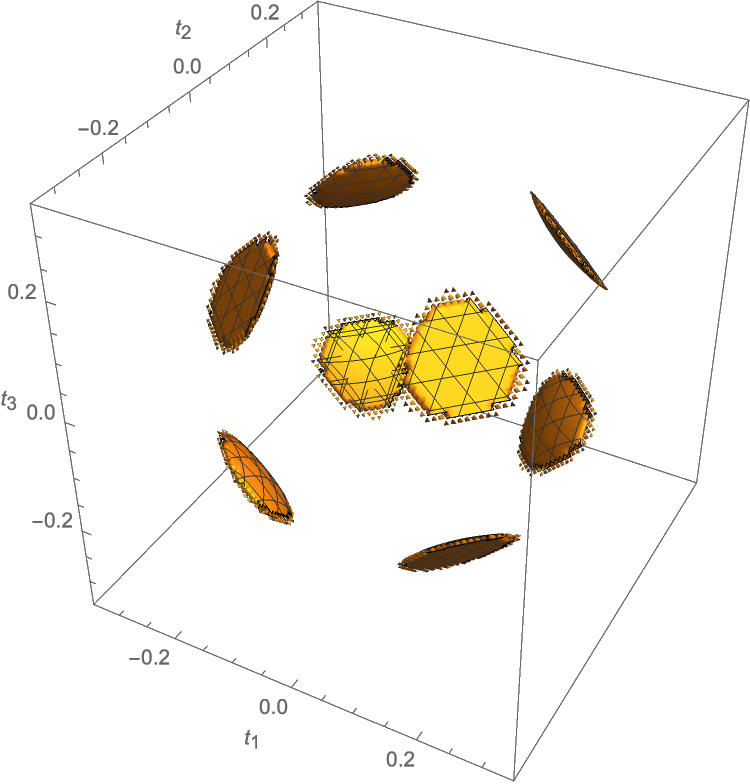}
    \caption{Bound-entangled archipelago of those two-ququart states (\ref{2ququarts}) satisfying the constraint $(t_1 t_2 t_3)^2 > \frac{1}{65536}$, but not  $(|t_1|+|t_2|+|t_3|)^2> \frac{1}{4} $. The associated probability is approximately 0.01403577037231.}
    \label{fig:TwoQuquartOneConstraint2}
\end{figure}

In Figure~\ref{fig:TwoQuquarttOneConstraint}, we show the bound-entangled  archipelago of those two-ququart states (\ref{2ququarts}) satisfying the constraint $(|t_1|+|t_2|+|t_3|)^2> \frac{1}{4} $, but not $(t_1 t_2 t_3)^2 > \frac{1}{65536}$. The associated probability is approximately 0.0175025342.
Reversing matters, in Figure~\ref{fig:TwoQuquartOneConstraint2}, we show the bound-entangled  archipelago of those two-ququart states  satisfying the constraint $(t_1 t_2 t_3)^2 > \frac{1}{65536}$, but not  $(|t_1|+|t_2|+|t_3|)^2> \frac{1}{4} $. The associated probability is quite negligible, that is approximately 0.01403577037231.

\begin{figure}
    \centering
    \includegraphics{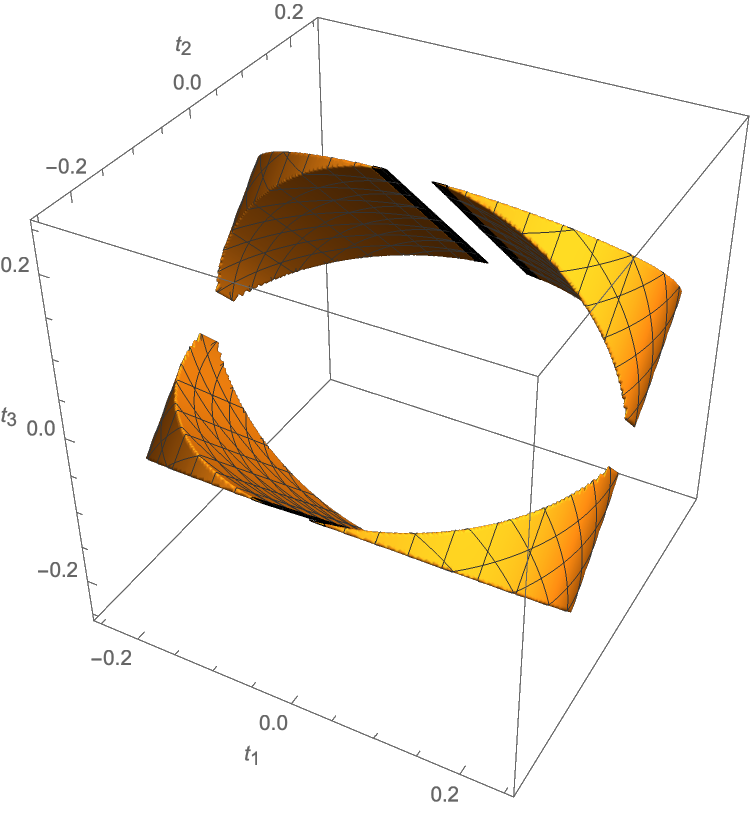}
    \caption{Those two-ququart states for the model (\ref{EQ4A}) that are entangled, but not bound, thus not PPT. Their probability is $\frac{3 \pi -8}{3 \pi } \approx 0.151174$.}
    \label{fig:Free2QQ}
\end{figure}

Let us now--somewhat briefly--study a second two-ququart model
\begin{equation} \label{2ququartsB}
\rho_{AB}^{(2)}=\frac{1}{2 \cdot 8} \textbf{4} \otimes \textbf{4} +\frac{1}{4} (t_1 \lambda_1 \otimes \lambda_1+t_2 \lambda_{9} \otimes \lambda_{9}+t_3 \lambda_{10} \otimes \lambda_{10}).
\end{equation}
The entanglement constraints are
\begin{equation} \label{EQ4A}
  (|t_1|+|t_2|+|t_3|)^2> \frac{1}{16} \hspace{.1in}  \mbox{or} \hspace{.1in} (t_1 t_2 t_3)^2 > 2^{-12} \cdot 3^{-6} = \frac{1}{2985984} \approx 3.34898 \cdot 10^{-7}
\end{equation}
The PPT-probability is $\frac{8}{3 \pi} \approx 0.848826$. The total (bound and free) entanglement probability is $\frac{3 \pi -4}{3 \pi } \approx 0.575587$, while the bound entanglement probability is $\frac{4}{3 \pi} \approx 0.424413$. The entangled but not bound states are shown in Fig.~\ref{fig:Free2QQ}.
For the convenience of the reader, and since the qubit-ququart and two-ququart analyses in \cite{slater2020archipelagos} have now been called into question, let us again 
present the interesting analyses there, not similarly suspect.

There, we  ``downgraded'' the Li-Qiao qubit-ququart model to  simply a two-qubit one,
\begin{equation} \label{rhoABtwoqubit}
\rho_{AB}^{(3)}=\frac{1}{2 \cdot 2} \textbf{1} \otimes \textbf{1} +\frac{1}{4} (t_1 \sigma_1 \otimes \sigma_1+t_2 \sigma_2 \otimes \sigma_{13}+t_3 \sigma_3 \otimes \sigma_3),
\end{equation}
while employing the entanglement constraints (again consistent with the leading-minors analysis),
\begin{equation} \label{eq8}
  (|t_1|+|t_2|+|t_3|)^2>1 \hspace{.1in}  \mbox{or} \hspace{.1in} (t_1 t_2 t_3)^2 > \Big(\frac{1}{27}\Big)^2.
\end{equation}
Then, we obtained a number of interesting results. Firstly, now only one-half of the physically possible states had positive partial transposes.

\begin{figure}
    \centering
    \includegraphics{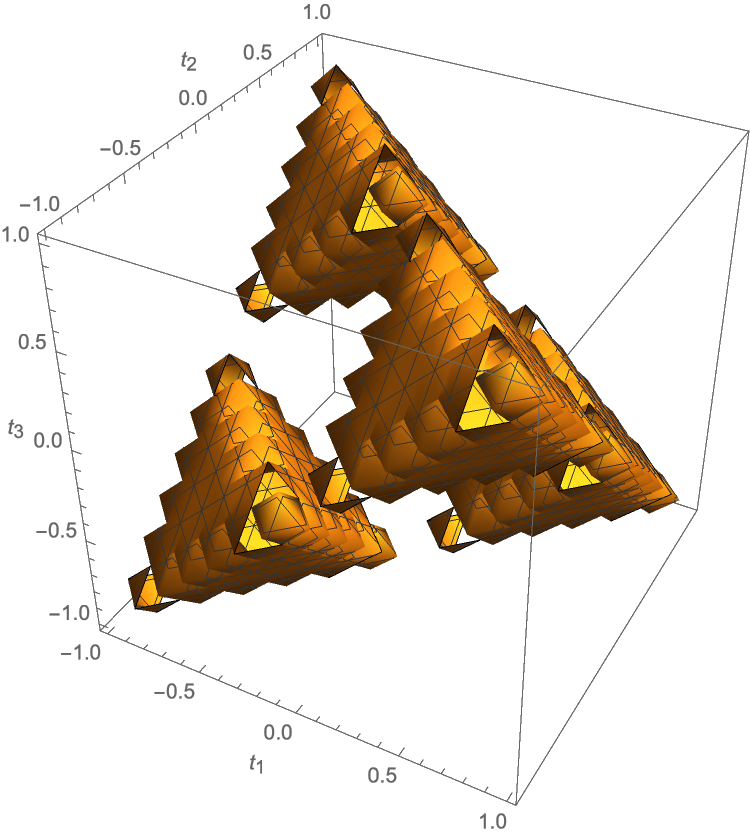}
    \caption{Archipelago of (non-bound/free) entangled two-qubit states for the set of states given by (\ref{rhoABtwoqubit}). The total probability is $\frac{1}{2}$.}
    \label{fig:TwoQubitsLiQiao}
\end{figure}

\begin{figure}
    \centering
    \includegraphics{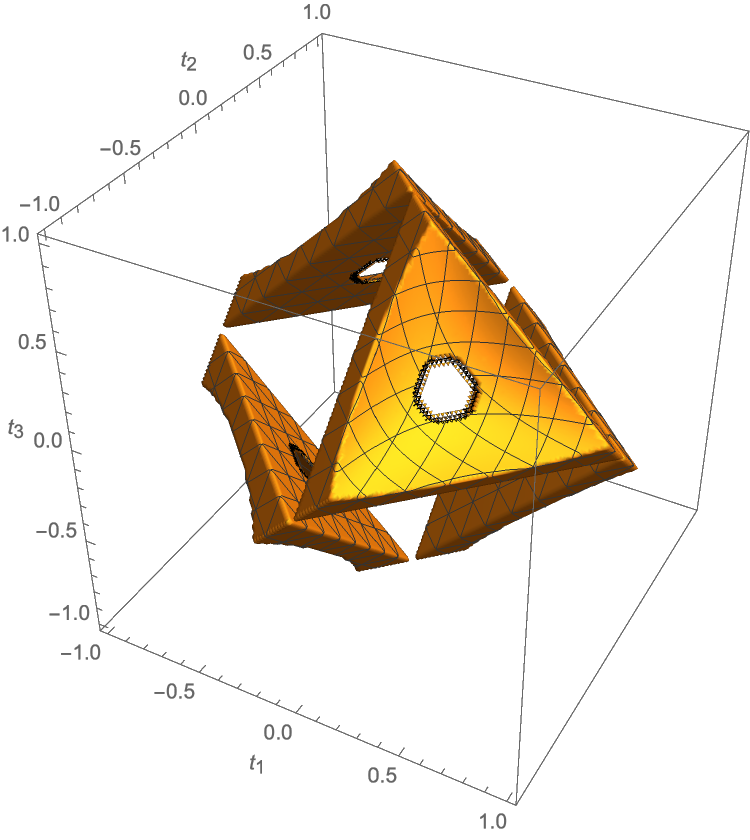}
    \caption{Those two-qubit states which satisfy the $(|t_1|+|t_2|+|t_3|)^2>1 $ entanglement  constraint, but not the $(t_1 t_2 t_3)^2 > \Big(\frac{1}{27}\Big)^2$ one. The associated probability is $\frac{1}{2}-0.3911856 =0.108814$.}
    \label{fig:PartFree}
\end{figure}

Also, imposition of the single (additive) constraint  $(|t_1|+|t_2|+|t_3|)^2>1 $ revealed that the other (non-PPT) half 
of the states are all entangled, as expected. On the other hand, enforcement of the single (multiplicative) constraint revealed that  only 0.3911855600402 of these non-PPT states were entangled. The entangled states again formed an archipelago (Fig.~\ref{fig:TwoQubitsLiQiao}), also apparently ``jagged'' in nature, but now clearly not of a bound-entangled nature (given the two-qubit context).
Those two-qubit states which satisfy the $(|t_1|+|t_2|+|t_3|)^2>1 $ entanglement  constraint, but not the $(t_1 t_2 t_3)^2 > \Big(\frac{1}{27}\Big)^2$, one are displayed in Fig.~\ref{fig:PartFree}. The associated probability is $\frac{1}{2}-0.3911856 =0.108814$.

Continuing with our analyses, we have been able to determine that the appropriate (multiplicative) entanglement constraint to employ for the first member, 
\begin{equation} \label{firstmember}
\rho_1= \frac{1}{9} \textbf{1} \otimes \textbf{1} + \frac{1}{4}(t_1\lambda_{1}\otimes \lambda_{1} +t_2 \lambda_{2}\otimes \lambda_{2}+ t_3\lambda_{3}\otimes \lambda_{3} ) \end{equation}
of the pair of
two-qutrit (octahedral and tetrahedral) models of Li and Qiao \cite[sec. 2.3.2]{li2018separable} is
\begin{equation}
( t_1 t_2 t_3 )^2> \frac{2^{12}}{3^{18}}=\frac{4096}{387420489},  
\end{equation}
and for the second member, 
\begin{equation} \label{secondmember}
\rho_2= \frac{1}{9} \textbf{1} \otimes \textbf{1} + \frac{1}{4}(t_1\lambda_{1}\otimes \lambda_{1} + t_2\lambda_{2}\otimes \lambda_{4}+ t_3\lambda_{3}\otimes \lambda_{6})  \end{equation}
of the pair,
\begin{equation}
 (t_1 t_2 t_3 )^2> \frac{2^{12}}{3^{15} } =\frac{4096}{14348907}.
\end{equation}
(We achieved these results by maximizing the product $t_1 t_2 t_3$,
subject to the conditions that the parameterized target density matrix and its separable components  not lose their positive definiteness properties.)

For the first two-qutrit model (\ref{firstmember}), we remarkably found the exact same entanglement behavior/probabilities ($\frac{1}{2}$ and 0.3911855600402 and Fig.~\ref{fig:PartFree}) as we did in  the two-qubit analyses.
Also, we did  not find that the second two-qutrit model (\ref{secondmember}) evinced any entanglement at all--in accordance with the explicit assertion of Li and Qiao that the state ``is separable for all values of $t_i$,\ldots"

As an additional two-qutrit exercise, let us consider the model
\begin{equation} \label{addmember}
\rho_1= \frac{1}{9} \textbf{1} \otimes \textbf{1} + \frac{1}{4}(t_1\lambda_{2}\otimes \lambda_{2} +t_2 \lambda_{4}\otimes \lambda_{4}+ t_3\lambda_{6}\otimes \lambda_{6} ) .
\end{equation}
The associated PPT probability is $\frac{1}{2}+\frac{2}{\pi ^2} \approx 0.702642$. 
The pair of entanglement constraints now takes the form
\begin{equation} \label{EntanglementAddendum}
  (|t_1|+|t_2|+|t_3|)^2> \frac{16}{81} \hspace{.1in}  \mbox{or} \hspace{.1in} (t_1 t_2 t_3)^2 > \frac{2^{12}}{3^{18}} = \frac{4096}{387420489} \approx 0.00001057249,
\end{equation}
The probability that a state (\ref{addmember}) satisfies the multiplicative constraint is 
0.490454, while the probability that it satisfies the additive constraint is $1-\frac{8}{3 \pi ^2} \approx 0.72981$. The corresponding bound-entanglement probabilities are 0.205794 and $\frac{1}{2}-\frac{2}{3 \pi ^2}  \approx 0.432453$.
The entirety of entanglement probability is 0.748599, while the entirety of 
bound-entangled probability is 0.43549.

\begin{figure}
    \centering
    \includegraphics{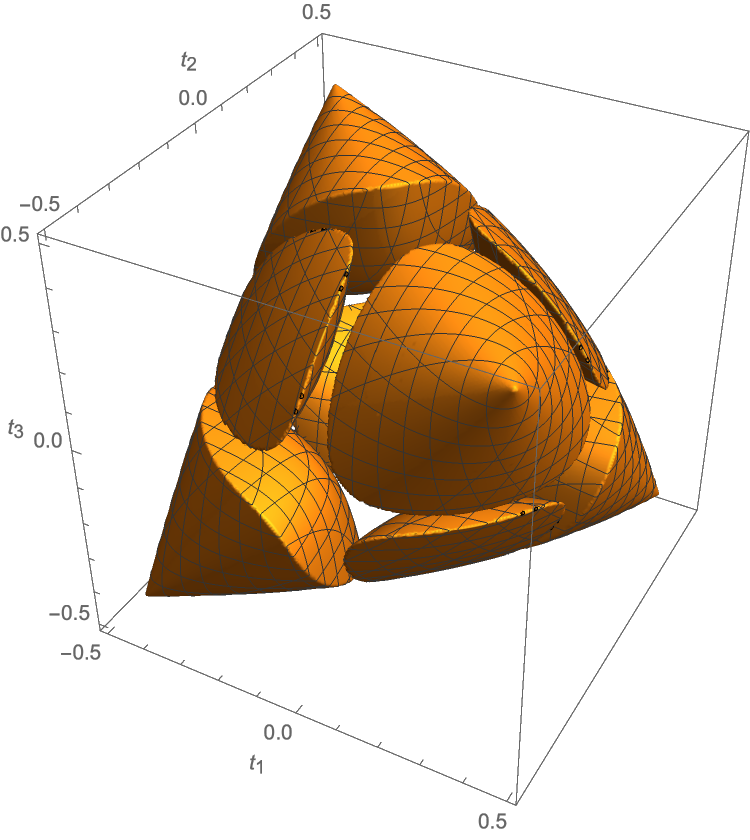}
    \caption{Those two-qutrit states (\ref{addmember}) that are entangled (free or bound) and satisfy both entanglement constraints (\ref{EntanglementAddendum}). The associated entanglement probability is 0.490454.}
    \label{fig:AddendumFreeBoundBoth}
\end{figure}

\begin{figure}
    \centering
    \includegraphics{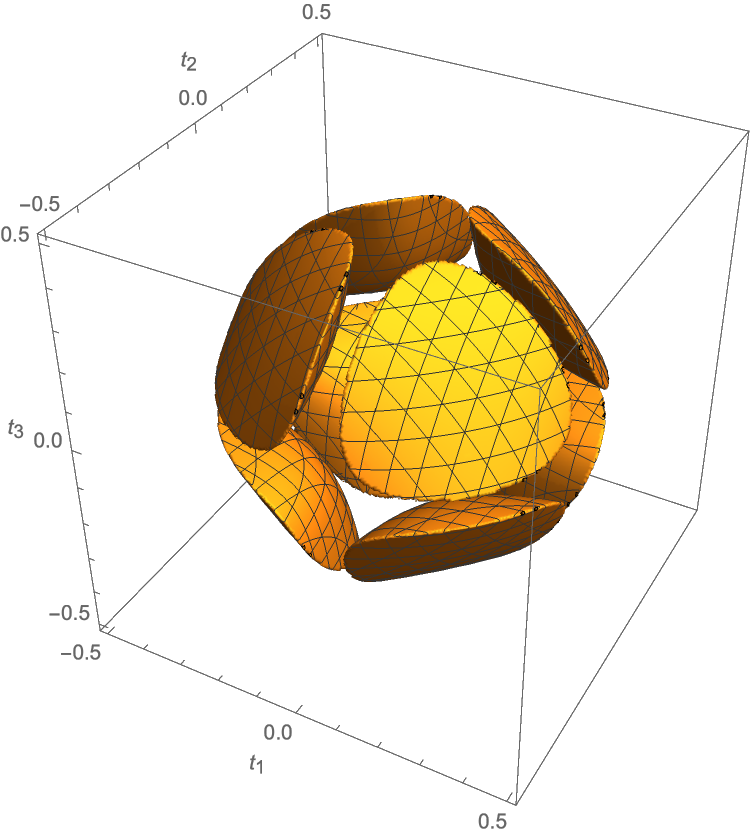}
    \caption{Those two-qutrit states (\ref{addmember}) that are bound and satisfy both entanglement constraints (\ref{EntanglementAddendum}). The associated bound entanglement probability is 0.205794.}
    \label{fig:AddendumBoundBoth}
\end{figure}

In Fig.~\ref{fig:AddendumFreeBoundBoth}, we show those (free or bound) entangled states satisfying both entanglement constraints.
On the other hand, in Fig.~\ref{fig:AddendumBoundBoth}, we show only bound entangled states satisfying both entanglement constraints.

To further pursue these general lines of investigation following the approach of Li and Qiao, we searched for qutrit-ququart models with non-positive-partial-transpose states. One that emerged took the form
\begin{equation} \label{qutritququart}
\rho= \frac{1}{12} \textbf{1} \otimes \textbf{1} + \frac{1}{4}(t_1\lambda_{4}\otimes \kappa_{1} + t_2\lambda_{6}\otimes \kappa_{6}+ t_3\lambda_{7}\otimes \kappa_{10}) ,    
\end{equation}
with the $\lambda$'s being as before the SU(3) generators and the $\kappa$'s now being the SU(4) generators.
The associated PPT-probability is $\frac{1}{2}+\frac{2}{\pi ^2}  \approx 0.848826$.
The relevant entanglement constraints are now 
\begin{equation} \label{qubitqutritentanglement}
  (|t_1|+|t_2|+|t_3|)^2>\frac{1}{9} \hspace{.1in}  \mbox{or} \hspace{.1in} (t_1 t_2 t_3)^2 > \frac{1}{531441}=3^{-12}.
\end{equation}
The entire (bound and free) entanglement probability based on the union of these two constraints is 
$\frac{3 \pi -4}{3 \pi } \approx 0.575587$, while the bound component is $\frac{4}{3 \pi} \approx 0.424413$. In fact, the first constraint fully dominates the second one. That is, there are no states entangled in terms of the second constraint that are not entangled in terms of the first. If we employ just the second constraint, then the corresponding entanglement probabilities are 0.304652 and 0.1706.

For the further (now PPT) qutrit-ququart model,
\begin{equation} \label{qutritququart2}
\rho_2= \frac{1}{12} \textbf{1} \otimes \textbf{1} + \frac{1}{4}(t_1\lambda_{2}\otimes \kappa_{1} + t_2\lambda_{3}\otimes \kappa_{3}+ t_3\lambda_{5}\otimes \kappa_{13}) ,    
\end{equation}
we have found  entanglement constraints of the form
\begin{equation} \label{qubitqutritentanglement2}
(|t_1|+|t_2|+|t_3|)^2>\frac{1}{9} \hspace{.1in}  \mbox{or} \hspace{.1in}  (t_1 t_2 t_3)^2 > \frac{411+41 \sqrt{41}}{123018750} \approx 5.4750035 \cdot 10^{-6}.
\end{equation}
where $123018750 =2 \cdot 3^9 \cdot 5^5$.  The associated bound-entanglement probabilities yielded by enforcement of the two constraints individually are  0.639747 and 0.185841, respectively. The first constraint fully dominates the second.

Following and building upon the work of Li and Qiao, all the analyses 
reported above have involved the {\it three} parameters $t_1, t_2, t_3$, thus, lending results to immediate visualization. In higher-dimensional studies, one would have to resort to cross-sectional examinations, such as Figs. 22 and 23 in \cite{slater2019bound}, based on the ({\it four} parameter) two-ququart  Hiesmayr-L{\"o}ffler ``magic simplex" model \cite{hiesmayr2014mutually}.

Of course, visualizations are possible in lower (two) dimensions, as well. In fact, we examined the two-qutrit (PPT) model
\begin{equation} \label{submember}
\rho= \frac{1}{9} \textbf{1} \otimes \textbf{1} + \frac{1}{4}(t_1 \lambda_{1} \otimes \lambda_{1} +t_2 \lambda_{4} \otimes \lambda_{4} ) .
\end{equation}
In doing so, in adopting the primary three-parameter Li-Qiao framework to a two-parameter one, we followed their prescriptions regarding the choice of orthogonal matrices $Q$, following eq. (23) in \cite{li2018necessary}. Such matrices are of dimension $(l+1) \times (l+1)$, where $l$ is the number of parameters. The last row of $Q$ contains non-negative entries. In particular, for the three subsequent (two-qutrit, two-ququart and qutrit-ququart) two-parameter analyses, we employed
\begin{equation}
 Q=\left(
\begin{array}{ccc}
 \frac{1}{\sqrt{6}} & -\sqrt{\frac{2}{3}} & \frac{1}{\sqrt{6}} \\
 \frac{1}{\sqrt{2}} & 0 & -\frac{1}{\sqrt{2}} \\
 \frac{1}{\sqrt{3}} & \frac{1}{\sqrt{3}} & \frac{1}{\sqrt{3}} \\
\end{array}
\right).   
\end{equation}

\begin{figure}
    \centering
    \includegraphics{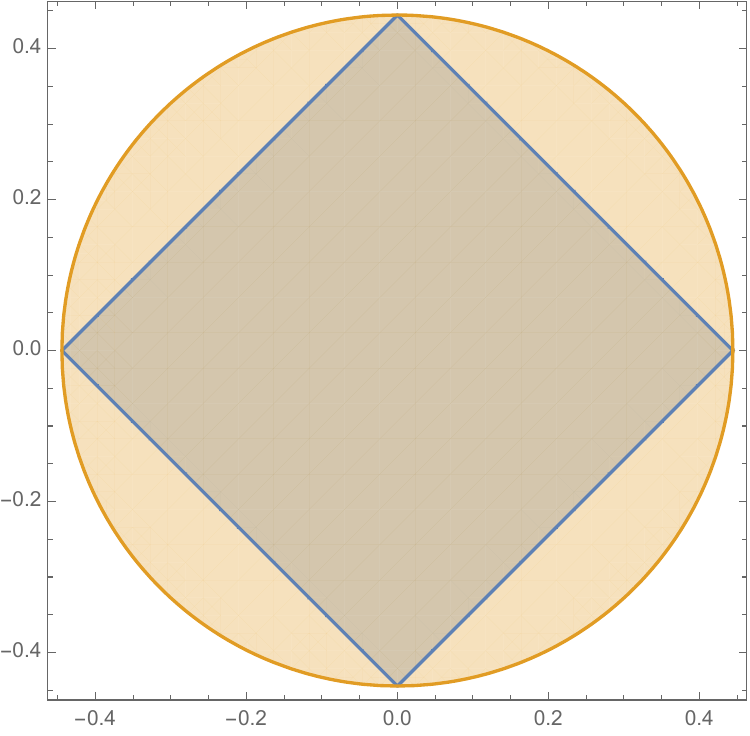}
    \caption{The circle comprises the possible states of the two-parameter two-qutrit model (\ref{submember}), while the inscribed square constitutes the {\it unentangled} states.
    The archipelago--of probability $\frac{\pi -2}{\pi } \approx 0.36338$--of states lying outside the square comprises the bound-entangled state.}
    \label{fig:TwoDimLiQiaoStates}
\end{figure}

\begin{figure}
    \centering
    \includegraphics{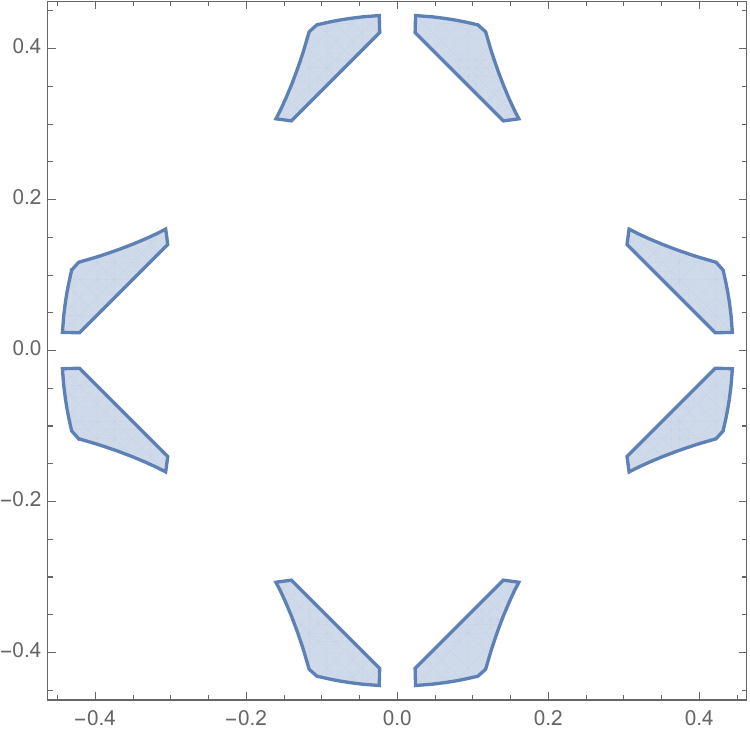}
    \caption{Those bound-entangled states of the  two-parameter two-qutrit model (\ref{submember}) that are revealed by the dominant constraint $(|t_1|+|t_2|)^2>\frac{16}{81} $, but not by the subdominant constraint 
 $ (t_1 t_2)^2 > \frac{16}{6561}$. The archipelago of bound-entangled states shown is of probability $\frac{-6+\pi +3 \cosh ^{-1}(2)}{3 \pi } \approx 0.115914$.}
    \label{fig:Subdominant}
\end{figure}

The entanglement constraints for the two-parameter model (\ref{submember}) are of the form
\begin{equation} \label{minorentanglement2}
(|t_1|+|t_2|)^2>\frac{16}{81} \hspace{.1in}  \mbox{or} \hspace{.1in}  (t_1 t_2)^2 > \frac{16}{6561}.
\end{equation}
The set of possible states of area $\frac{16 \pi}{81} \approx 0.620562$ is the circle
$16 -81 t_1^2-81 t_2^2 \geq 0$ of radius $\frac{4}{9}$. The set of {\it unentangled} states is the inscribed square with vertices at $(\pm \frac{4}{9},0)$ and $(0, \pm \frac{4}{9})$.
This is shown in Fig.~\ref{fig:TwoDimLiQiaoStates}.
 The bound-entangled states, lying outside the inscribed square, are of probability  $\frac{\pi -2}{\pi } \approx 0.36338$. The constraint $(|t_1|+|t_2|)^2>\frac{16}{81} $ fully dominates the constraint $ (t_1 t_2)^2 > \frac{16}{6561}$ (which itself yields $\frac{2}{3}-\frac{\cosh ^{-1}(2)}{\pi } \approx 0.247466$). Those bound-entangled states that are yielded by the dominant constraint $(|t_1|+|t_2|)^2>\frac{16}{81} $, but not by the subdominant constraint 
 $ (t_1 t_2)^2 > \frac{16}{6561}$ are displayed in Fig.~\ref{fig:Subdominant}. They are of 
 probability $\frac{-6+\pi +3 \cosh ^{-1}(2)}{3 \pi } \approx 0.115914$.

Let us move on, still within the modified {\it two}-parameter Li-Qiao framework to the (PPT) two-ququart model
\begin{equation} \label{submemberququart}
\rho= \frac{1}{16} \textbf{1} \otimes \textbf{1} + \frac{1}{4}(t_1 \kappa_{7} \otimes \kappa_{7} +t_2 \kappa_{9} \otimes \kappa_{9} ) ,
\end{equation}
where the $\kappa$'s as in (\ref{qutritququart}) and (\ref{qutritququart2}) represent the $SU(4)$ generators with their standard ordering.
The entanglement constraints are of the form
\begin{equation} \label{minorentanglement3}
(|t_1|+|t_2|)^2>\frac{49}{576} \hspace{.1in}  \mbox{or} \hspace{.1in}  (t_1 t_2)^2 > \frac{1}{2304}.
\end{equation}
We will find that although the first constraint does not fully dominate the second, it nearly does--except for an archipelago of four regions accounting for only $\frac{7-24 \log \left(\frac{4}{3}\right)}{8 \left(4-3 \log
   \left(\frac{4}{3}\right)\right)} \approx 0.381063 \%$ of the total bound-entangled probability of $\frac{1}{9} \left(4-3 \log \left(\frac{4}{3}\right)\right)\approx 0.34855$.

\begin{figure}
    \centering
    \includegraphics{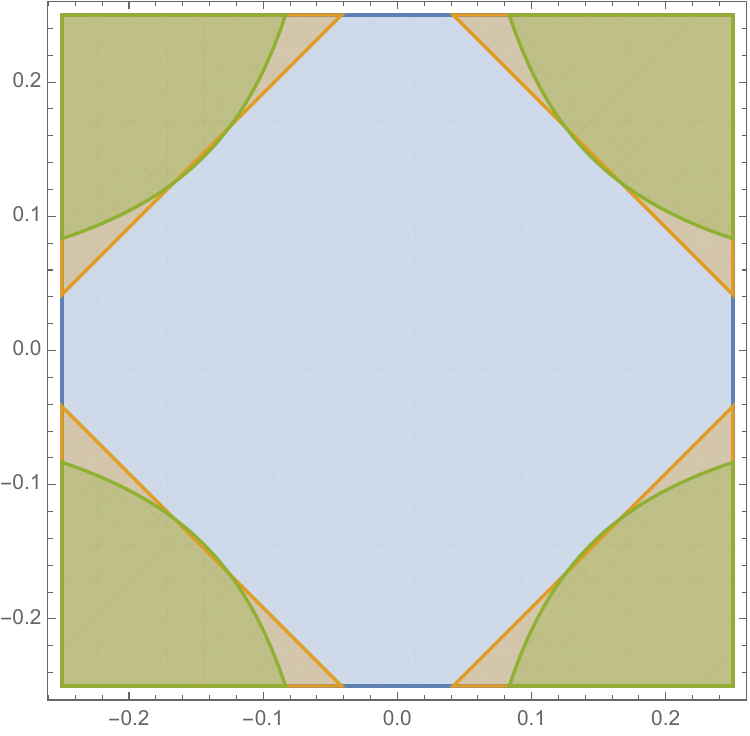}
    \caption{Full state space--the square--and the total bound-entangled region--the four corner triangles--obtained for the two-parameter two-ququart model (\ref{submemberququart}). The four curve-bounded corner subregions of the triangles comprise those (highly) bound-entangled states satisfying both entanglement constraints (\ref{minorentanglement3}).}
    \label{fig:twoparameterququart}
\end{figure}

In Fig.~\ref{fig:twoparameterququart} we show the square with vertices at $(\pm \frac{1}{4},\pm \frac{1}{4} )$, comprising the set of possible states. The four corner triangles of it comprise the bound-entangled states of the noted probability $\frac{1}{9} \left(4-3 \log \left(\frac{4}{3}\right)\right)\approx 0.34855$. The eight-sided region consists of the complementary separable states. The constraint $(|t_1|+|t_2|)^2>\frac{49}{576}$ accounts for $\frac{25}{72} \approx 0.347222$, and the constraint $(t_1 t_2)^2 > \frac{1}{2304}$ for $\frac{1}{3} (2-\log (3)) \approx 0.300463$. The bound-entangled probability attributable to the $(t_1 t_2)^2 > \frac{1}{2304}$ constraint, but not the other is only $\frac{1}{72} \left(7-24 \log \left(\frac{4}{3}\right)\right) \approx 0.0013282$. On the other hand, the bound-entangled probability attributable to the $(|t_1|+|t_2|)^2>\frac{49}{576}$ constraint, but not the other is  $\frac{2}{9} \left(\log \left(\frac{27}{8}\right)-1\right) \approx 0.0480878$.
(We also considered several further two-parameter scenarios--these of a {\it hybrid} qutrit-ququart character. They largely yielded diagrams of a rather similar nature to Fig.~\ref{fig:twoparameterququart}.)

It now seems possible to rather readily extend the Li-Qiao framework to further high-dimensional bipartite systems--{\it e. g.} qutrit-ququart, qubit-ququint,\ldots other than the specific ones studied above. Of immediate interest for all such systems is the question of to what extent they have positive partial transposes. Then, issues of bound and free entanglement can be addressed.

Let us also raise the question of whether or not the Hiesmayr-L{\"o}ffler ``magic simplices"  \cite{hiesmayr2014mutually} and/or the generalized Horodecki states \cite{horodecki1998mixed} can be studied--through reparameterizations--within the Li-Qiao framework, with consequent answers as to the associated {\it total} bound entanglement probabilities. Possibly, then, the new archipelagos might not evince the strong jaggedness previously observed \cite{slater2019bound}, along the lines of those observed above here. Jaggedness, then, being a feature of incompleteness/non-totality.

The two-qutrit Hiesmayr-L{\"o}ffler `magic simplex model is, expressible, we have found as
\begin{equation} \label{HLCorrelation}
\rho_{HL}= \frac{1}{9} \textbf{1} \otimes \textbf{1} + \frac{1}{4} \Big(t_9 \lambda_{3} \otimes \lambda_{8} +t_{10} \lambda_{8} \otimes \lambda_{3} +\Sigma_{i=1}^8 t_i \lambda_{i} \otimes \lambda_{i})\Big) .    
\end{equation}
(Interestingly, in the three-dimensional matrix [Gell-mann] representation of $SU(3)$, the Cartan subalgebra is the set of linear combinations (with real coefficients) of the two matrices 
$\lambda_3$ and $\lambda_8$, which commute with each other.)
Here, $t_1=t_4=t_6=\frac{2}{3} \left(Q_1-Q_3\right), t_2=t_5=-\frac{2}{3} \left(Q_1-Q_3\right), t_3=t_8 =-(1/3)+Q_1+2 Q_3$. Further, $t_9=\frac{Q_1+6 Q_2+2 Q_3-1}{\sqrt{3}}$ and $t_{10}=-\frac{Q_1+6 Q_2+2 Q_3-1}{\sqrt{3}}$.

Six of the eight singular values of the correlation matrix of (\ref{HLCorrelation}) are $u=\frac{2}{3} \sqrt{\left(Q_1-Q_3\right){}^2}$ and the remaining two are $v=\frac{2}{3} \sqrt{-9 Q_2-6 Q_3+3 \left(Q_1^2+\left(3 Q_2+4 Q_3-1\right) Q_1+9 Q_2^2+4
   Q_3^2+6 Q_2 Q_3\right)+1}$.
   
  Numerical analyses appear to strongly indicate that one of the corresponding Li-Qiao entanglement constraints is 
   $(2 \left| u\right| +6 \left| v\right| )^2 > \frac{1}{144}$.

Gabuldin and Mandilara concluded that the particular bound-entangled states they found in certain analyses of theirs had ``negligible volume and that these form tiny `islands' sporadically distributed over the surface of the polytope of separable states" \cite{gabdulin2019investigating}. In a continuous variable study \cite{diguglielmo2011experimental}, ``the tiny regions in parameter space where bound entanglement does exist'' were noted.

Let us note the recent posting of a paper entitled "Entanglement islands in higher dimensions" \cite{almheiri2019entanglement}, concerned with the famous information paradox. 
The authors conclude: ``Islands appear in entanglement wedge of the Hawking radiation at late times and this stops the indefinite growth of von Neumann entropy, giving an answer consistent with unitarity and a finite density of states."

We further observed that the matrix $Q \in SO(4)$,
\begin{equation}
 Q=\frac{1}{2} \left(
\begin{array}{cccc}
 1 & -1 & -1 & 1 \\
 -1 & -1 & 1 & 1 \\
 -1 & 1 & -1 & 1 \\
 1 & 1 & 1 & 1 \\
\end{array}
\right)
\end{equation}
employed by Li and Qiao \cite[eq. (62)]{li2018separable}
is a $4 \times 4$ Hadamard matrix \cite{horadam2012hadamard}. So, we investigated the possibility that by employing the $8 \times 8$
Hadamard matrix
\begin{equation}
\tilde{Q} =\frac{1}{\sqrt{8}} \left(
\begin{array}{cccccccc}
 -1 & 1 & 1 & -1 & 1 & -1 & -1 & 1 \\
 1 & 1 & -1 & -1 & -1 & -1 & 1 & 1 \\
 1 & -1 & 1 & -1 & -1 & 1 & -1 & 1 \\
 -1 & -1 & -1 & -1 & 1 & 1 & 1 & 1 \\
 1 & -1 & -1 & 1 & 1 & -1 & -1 & 1 \\
 -1 & -1 & 1 & 1 & -1 & -1 & 1 & 1 \\
 -1 & 1 & -1 & 1 & -1 & 1 & -1 & 1 \\
 1 & 1 & 1 & 1 & 1 & 1 & 1 & 1 \\
\end{array}
\right)    
\end{equation}
we might extend the Li-Qiao framework from a $3=4-1$-dimensional one to an $7=8-1$-dimensional one. Accordingly--as one of eight possible options--we set up the two-qutrit model
\begin{equation} \label{addmemberHadamard}
\rho_1= \frac{1}{9} \textbf{1} \otimes \textbf{1} + \frac{1}{4} \Sigma_{i=1}^7 t_i \lambda_{i}\otimes \lambda_{i} ,
\end{equation}
where  the $\lambda$'s are the $SU(3)$ generators. ($\lambda_8$ is the single one not employed.) For this model, we obtained a PPT-probability of 0.662799194015. Our attempts to obtain the corresponding entanglement constraints and entanglement probabilities have so far not yielded numerical results in which we have sufficient confidence to report.

\begin{acknowledgements}
This research was supported by the National Science Foundation under Grant No. NSF PHY-1748958.
\end{acknowledgements}

\bibliography{main}

\end{document}